\documentclass[[12pt]{article}
\usepackage{verbatim}
\usepackage[english]{babel}
\usepackage{pifont}
\usepackage{amsmath}
\usepackage{graphicx}
\usepackage[letterpaper, margin=1in]{geometry}
\usepackage{natbib}

\usepackage{mathptmx}      
\usepackage{latexsym}
\usepackage[affil-it]{authblk}

\begin{document}

\title{Long-run evolution of the global economy -- Part~2: Hindcasts of innovation and growth}

\author{Timothy~J.~Garrett}
\date{}

\affil{Department of Atmospheric Sciences, University of Utah, Salt Lake City, UT, USA, {tim.garrett@utah.edu}}





\maketitle

\begin{abstract}
Long-range climate forecasts use integrated assessment models to link the
global economy to greenhouse gas emissions. This paper evaluates an
alternative economic framework outlined in part 1 of this study \citep{GarrettEF2014} that
approaches the global economy using purely physical principles rather than
explicitly resolved societal dynamics. If this model is initialized with
economic data from the 1950s, it yields hindcasts for how fast global
economic production and energy consumption grew between 2000 and 2010 
with skill scores $>$\,90\,\% relative to a model of persistence in trends. The
model appears to attain high skill partly because there was a strong impulse
of discovery of fossil fuel energy reserves in the mid-twentieth century that
helped civilization to grow rapidly as a deterministic physical response.
Forecasting the coming century may prove more of a challenge because the
effect of the energy impulse appears to have nearly run its course.
Nonetheless, an understanding of the external forces that drive civilization
may help development of constrained futures for the coupled evolution of
civilization and climate during the Anthropocene.
\end{abstract}

\section{Introduction}

Climate simulations require as input future scenarios for greenhouse gas
emissions from integrated assessment models~(IAMs). IAMs are designed to
explore how best to optimize societal well-being while mitigating climate
change. The calculations of human behaviors are made on a regional and
sectoral basis and can be quite complex, possibly with hundreds of equations
to account for the interplay between human decisions, technological change,
and economic growth \citep{moss2010,IPCC_WG32014}.

Periodically, model scenarios are updated to account for observed emissions
trajectories. For example, it has been noted that the global carbon dioxide
(CO$_{2}$) emission rate has not only grown along a
``business-as-usual''~(BAU) trajectory but has in fact slightly exceeded it
\citep{Raupach2007,Peters2013}, in spite of a series of international accords
aimed at achieving the opposite \citep{Nordhaus2010}.

What stability in emissions growth might suggest is that the human system has
inertia, much like physical systems. Current variability reflects an
accumulation of prior events, so persistent forces from the past tend to have
the greatest influence on the present. Such large-scale trends tend to
continue to persist into the future because they are the least responsive to
current small-scale rapid forces that become diluted in the history of
actions that preceded them \citep{Hasselman1976}. It may be that it is
difficult to wean ourselves from fossil fuels today because we have spent at
least a century accumulating a large global infrastructure for their
consumption. It is not that current efforts to move civilization towards
renewables cannot change this trajectory of carbon dependency but rather
that it will take considerable effort and time.

\begin{figure}[t]\centering
\includegraphics[width=14cm]{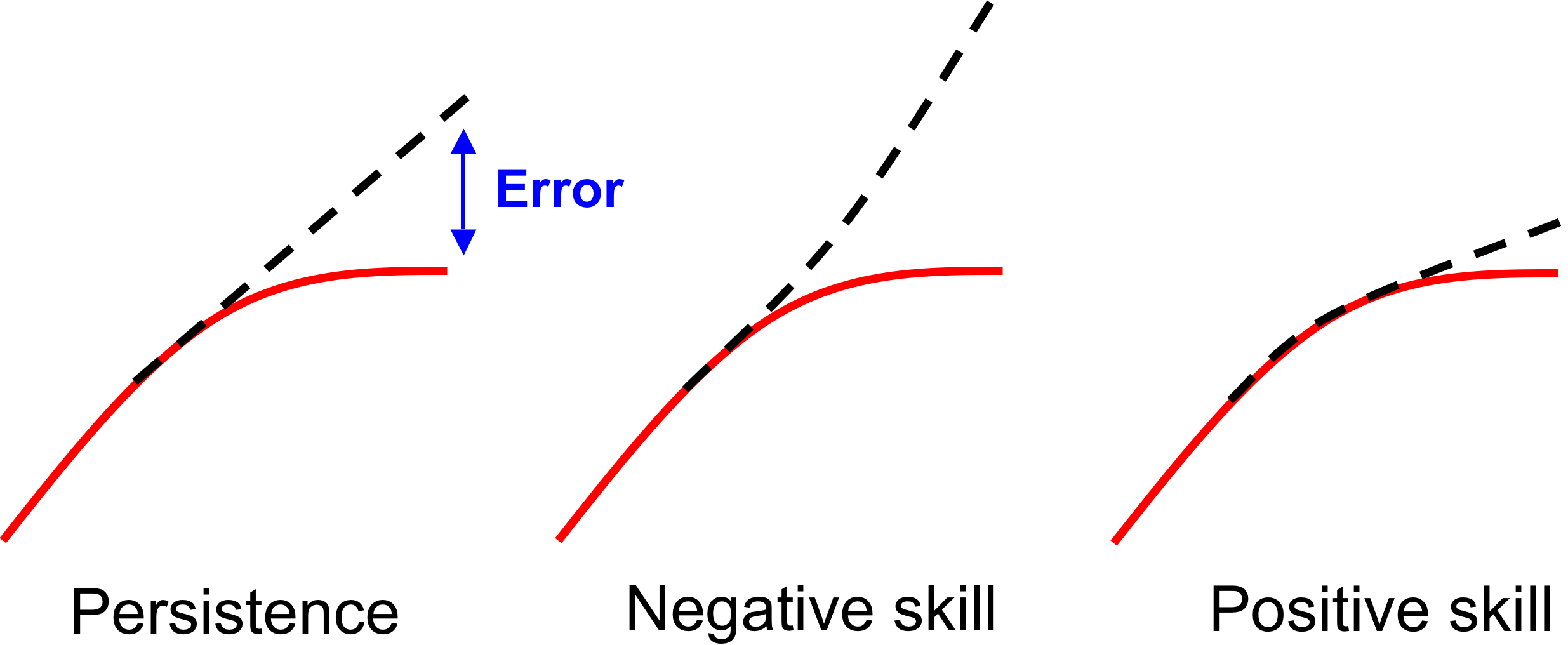}
\caption{Positive skill in forecasts (black line) requires doing better than
persistence in predicting future evolution of a quantity (red line).\label{fig:skill}}
\end{figure}

Inertia offers plausibility to a BAU trajectory, particularly
for something as highly integrated in the space and time as CO$_{2}$
emissions by civilization as a whole. Still, assuming persistence in trends
is something that should only be taken so far. By analogy to meteorological
forecasts, it is reasonable to assume that clearing skies will lead to
a sunny day. However, prognostic weather models are based on fundamental
physical principles that tell us that it cannot keep getting sunnier. Even
a very simple set of equations dictates that at some point a front will pass,
clouds will form, and a high-pressure system will decay. It is by getting the
underlying physics right that we are able to achieve some level of
\textit{positive skill} in any forecast attempt (Fig.~\ref{fig:skill}).

The macroeconomic components of IAMs do not offer true forecasts that can be
assigned a skill score. Rather they reflect expert opinions \citep{moss2010}
and are mostly unconstrained by external physical forces since they are
policy-driven and mathematically constructed so as to allow for an extremely
broad range of possible futures \citep{Pindyck2013}. They consider labor,
physical capital, and human inspired technological change to be the motive
forces for economic production and growth. The focus is on individuals,
nations, and economic sectors. The model equations describe how physical
capital and human prosperity grow with time, and how energy choices tie in
with greenhouse gas emissions \citep{Solow1956,Nordhaus2013DICE}.

Part~1 of this study (\citealp{GarrettEF2014}, hereafter referred to as Part~1)
described a second, more deterministic approach. CO$_{2}$ is considered
to be long-lived and well mixed in the atmosphere, so the magnitude of
greenhouse forcing is almost entirely unrelated to the national origin of
anthropogenic emissions. Then, civilization can be described as a whole, one
where small-scale details at personal, regional, or sectoral levels are not
treated explicitly. The only quantity in the model that needs to be resolved
is an aggregated global economy that is inclusive of all civilization
elements, including human and physical capital combined.

As an alternative to IAMs, this new approach offers a means for integrating
human systems and physical systems under a common framework, one where the
governing equations are consistently derived from first thermodynamic
principles. Much like the primitive equations of a prognostic weather model,
global economic growth is expressed as a non-equilibrium response to external
gradients driving energy dissipation and material flows. There is no explicit
role for human decisions; physics does not readily allow for mathematical
expressions of policy. Rather, economic innovation and growth is treated
primarily as a geophysical phenomenon, in other words the totality of
civilization is expressed as an emergent response to available reserves of
raw materials and energy supplies.

Whatever the approach that is applied, it is important that any societal
model be evaluated for performance. Weather, climate, and financial models
are regularly evaluated through hindcasts or backtesting. Economic models
that simulate the long-run development of humanity need not be an exception.
A good model, even one that includes policy, should be able to reproduce
current events with positive skill starting at a point some decades in the
past. To beat the zero-skill hindcast of persistence, the model would invoke
fits to concurrent trends to the minimum extent possible.

In the period following World War II, an economic ``front'' passed that
propelled civilization towards unprecedented levels of prosperity and, by
proxy, greenhouse gas emissions. This paper examines whether the theoretical
model introduced in Part~1 can explain the evolution of this front. Section~2
outlines the philosophical and thermodynamic basis for describing economic
evolution with physics. Section~3 evaluates this model from hindcasts.
Sections~4 and~5 discuss and summarize the results.

\section{Forces for economic growth}

\subsection{The relationship of energy dissipation to human wealth}

A formal framework for the non-equilibrium thermodynamics of civilization
growth was laid out in Part~1 \citep{GarrettEF2014}. The basis for a model
linking economics to physics is a fundamental identity that relates a
monetary expression for wealth to the rate at which civilization powers
itself with primary energy sources \citep{GarrettCO2_2009}. We all have some
sense of the difference between civilization and its uncivilized surroundings
since we have farms, buildings, human population, vehicles, and communication
networks. As shown in Fig.~\ref{fig:reversible-irreversible} and discussed in
Appendix~A, this difference implies the existence of a gradient between
civilization and its environment. Gradients allow for irreversible
thermodynamic flows. A consumption and dissipation of potential energy by
civilization sustains internal reversible circulations within civilization
that characterize all its activities.

The hypothesis that was made in Part~1 is that civilization is effectively a
heat engine whose power can be represented in more human terms as economic
wealth. Absent any energy consumption, civilization would necessarily decay
towards an uncivilized, worthless equilibrium where the gradient ceased to
exist and all internal circulations stopped. Wealth is able to grow only when
net work is done to grow into the environment. Real economic production
occurs when raw materials can be incorporated into civilization's structure
faster than civilization decays (Fig.~\ref{fig:reversible-irreversible}).
Growth at a net positive rate expands civilization's interface with reserves
of energy. It enlarges civilization, creating new wealth and a greater
overall capacity of civilization to consume and sustain internal
circulations. Expressed as an integral, current economic wealth is the net
accumulation of past net physical work and real economic production.


In effect, there is no intrinsic wealth in and of itself. Rather, as
discussed in Part~1, wealth is built from a network of connections, from the
product of a material length density and an energetic potential. Connections
are what enable dissipative flows insofar as there exist potential energy
gradients to drive the flows. For civilization as a whole, wealth is
sustained by primary power consumption through the connections we have to
reserves of fossil, nuclear, and renewable energy sources. Within
civilization, the interpretation is that wealth is due to the connections
between and among ourselves and our ``physical capital'', and from the
circulations along transportation, telecommunications, and social networks.
All aspects of civilization, whether social or material, compete for globally
available potential energy. Financial expressions of any element's value
reflect the relative extent to which its connections enable civilization to
irreversibly consume potential energy in order to sustain the reversible
circulations of the global economy.

While energy consumption is required to economically produce and grow, a
generally much greater amount is required to sustain circulations within the
networks of connections that have accumulated from prior production in the
past. An analog is an adult human. People's bodies are also a network of
connections that have grown through childhood and adolescence. Far more of
current daily food consumption goes towards maintaining life than to any
extra weight gain. Similarly, value added to civilization through
construction of a house decades ago still contributes to value today, by
being part of a larger network that supports the daily rhythms of its inhabitants.

\begin{figure}[t]\centering
\includegraphics[width=14cm]{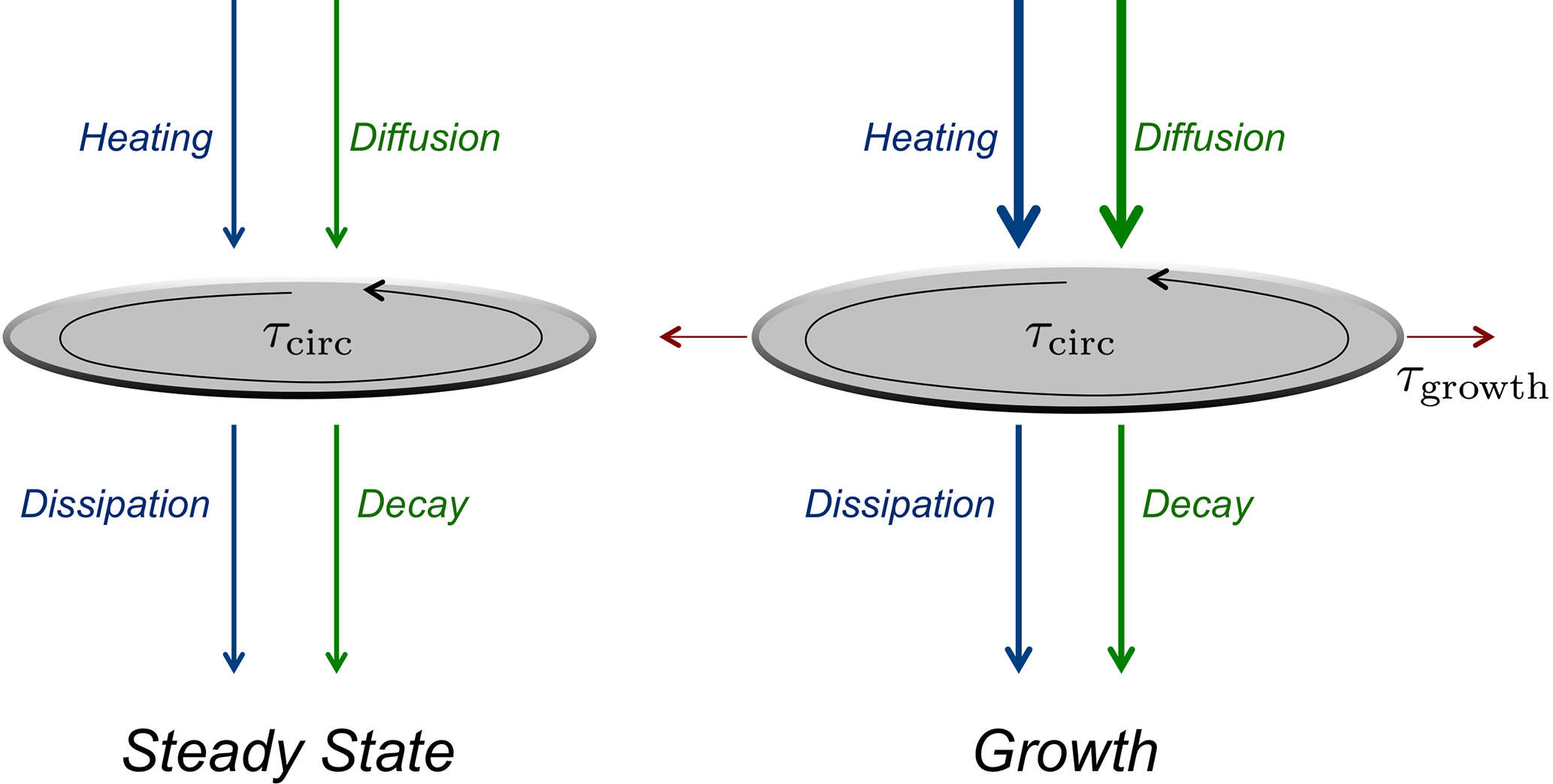}
\caption{Thermodynamic representation of an open system. Reversible
circulations within a system that lies along a constant potential have
a characteristic time $\tau_\textrm{circ}$. Circulations are sustained by
a dissipation of a potential energy source that heats the system. The system
maintains a steady state (left panel) because energetic (blue) and material (green)
flows enter and leave the system at the same rate. Where there is a positive
imbalance (right panel), the system grows irreversibly with timescale
$\tau_\textrm{growth}$\,$\gg$\,$\tau_\textrm{circ}$. See Appendix~A for details.\label{fig:reversible-irreversible}}
\end{figure}

The analytical formulation of this hypothesis is that instantaneous power
dissipation, or the rate of primary energy consumption~$a$ by all of
civilization (expressible in units of energy per time or power), is linked through a constant~$\lambda$
(expressible in units of power per currency) to civilization's inflation-adjusted economic value
(or civilization wealth) $C$ (expressible in units of currency). Wealth is defined as an
accumulation of the gross world product~(GWP) $Y$, adjusted for inflation at
market exchange rates~(MER) \citep{GarrettCO2_2009}. MER units are used
rather than purchasing power parity units~(PPP) since the focus is not on
short-term inequalities between people and nations but rather the sum of all
activities within the global economy with an eye to variability in the long run.
Thus,
\begin{align}
& a=\lambda{C}=\lambda\int\limits_{0}^{t}Y(t')\textrm{d}t'.
\label{eq:alambdaC}
\end{align}
Alternatively, and taking the derivative with respect to time, economic
production is a representation of the growth of wealth:
\begin{align}
& \frac{\textrm{d}C}{\textrm{d}t}=Y,
\label{eq:P-dadt}
\end{align}
where, since $a$\,$=$\,$\lambda{C}$, the production function is given by an increase
in the capacity to consume energy:
\begin{align}
& Y=\frac{1}{\lambda}\frac{\textrm{d}a}{\textrm{d}t}.
\label{eq:productionfunction}
\end{align}

\begin{figure}[t]\centering
\includegraphics[width=14cm]{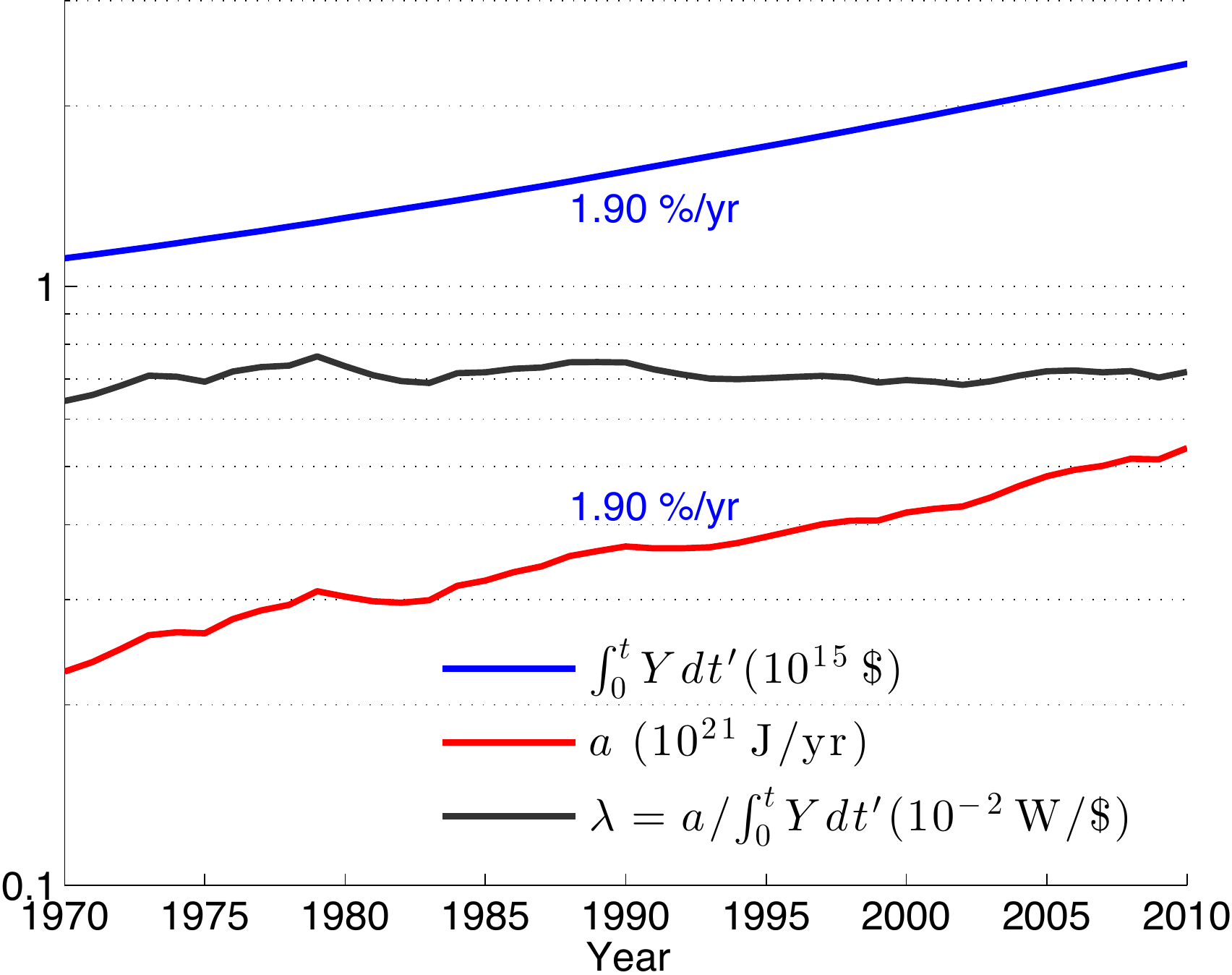}
\caption{Rates of global energy consumption~$a$, global wealth
$C$\,$=$\,$\int\limits_{0}^{t}\,Y(t')$\,d$t'$, and the ratio $\lambda$\,$=$\,$a/C$
since 1970. The average rates of growth~$\eta$ for $a$ and $C$ in percent per
year are shown for comparison. The average value of $\lambda$ is
7.1\,$\pm$\,0.1\,mW per year 2005 USD. Note the $y$~axis is a logarithmic scale. \label{fig:lambdaplot}}
\end{figure}

Crucially, Eq.~(\ref{eq:alambdaC}) is a hypothesis that can be tested
using available data. As described in greater detail in the Supporting
Information of Part~1, GWP estimates from \citet{Maddison2003} and the United
Nations \citep{UNstats} are used for historical estimates of $Y$. Estimates
of the global rate of primary energy consumption~$a$ are provided by the US
Department of Energy \citep{AER2011}. Expressing $a$ in units of watts, and
$Y$ in units of 2005~MER US dollars per second, then wealth has units of
2005~MER US dollars and the constant~$\lambda$ has units of watts per 2005 MER US
dollar. What was shown in Table~S2 of \citet{GarrettEF2014}, and in graphical
form in Fig.~\ref{fig:lambdaplot}, is that, for the period 1970 to 2010 for
which global statistics for power consumption are available, both $a$ and
$\int\limits_{0}^{t}\,Y(t)$\,d$t'$ have risen nearly in lockstep.
The mean value of $\lambda$ relating the two quantities is 7.1\,mW per
2005~US dollar. Even though the GWP more than tripled over this time period,
from year to year, the SD in the ratio $\lambda$\,$=$\,$a/C$ was just 1\,\%,
implying an uncertainty in the mean at the 95\,\% confidence level of
0.1\,mW per 2005~US dollar.

The constant~$\lambda$ is not derived from a correlation analysis (something
that has been erroneously claimed by others,
\citealp{Cullenward2010,Scher2010}), but instead it is obtained from the
observation that the ratio of $C$ to $a$ has not changed from year to year
even as $C$ and $a$ have. The observation is much like the basic expression
of quantum mechanics where it was initially assumed, and then confirmed with
measurements, that a photon's energy $E$ and its frequency $\nu$ are linked
through Planck's constant $h$. The empirical support for $\lambda$ or $h$
being effectively a constant stands on its own. But for the purposes of
understanding the physics, the quantities they relate are not correlated but
instead can be viewed as being interchangeable representations of the same thing.

The challenge might be to comprehend how a psychological construct like money
could be tied to a thermodynamic construct like power through a constant.
Economic value only goes so far as human judgement. Even with no one home and
all the utilities turned off, a house still maintains some worth for as long
as it can be perceived as being potentially useful by other active members of
the global economy.

The interpretation might be that physical
flows tie our brains to the global economy.
Brains process a wealth of information from our
environment using extraordinarily dense networks of axons
and dendrites; patterns of oscillatory
neuronal activity lead to the emergence of behavior and cognition;
powering this brain activity requires approximately 20\,\% of the daily caloric input to
the body as a whole \citep{Varela2001, Lennie2003, buzsaki2004}. Perhaps
dissipative neuronal circulations along brain networks reflect our collective perception of real
global economic wealth. They march to broader economic circulations
along global civilization networks that are sustained by a dissipation of oil, coal, and other primary
energy supplies. Eqs.~(\ref{eq:alambdaC}) to~(\ref{eq:productionfunction})
may seem unorthodox by traditional economic
standards, but there may be some basis for interpreting $\lambda$ as a type of
psychological constant that links the physics of human perception to the
thermodynamic flows that drive the global economy.

As a point of comparison, traditional economic growth models are divorced
from expressions of energy dissipation and physical flows, where wealth is
expressed in terms of a physical capital, or as a stock that has an intrinsic
value. New capital is produced using currently existing labor and capital, and
production levels have no explicit dependence on external physical
constraints \citep{Solow1956}. A more detailed outline and juxtaposition with
the model here is described in Appendix~B.

The field of macroeconomics is however making
steps towards creating links with physics, pointing out that, along with labor
and capital, energy must also be a factor of economic production
\citep{Lotka1922,Soddy1933,Odum1971,Georgescu-Roegen1993,Hua2004,AnnilaSalthe2009}.
Quantified links between physical and financial quantities often rely upon
a high observed correlation between national or sectoral economic production
and energy consumption \citep{Costanza1980,Cleveland1984,Brown2011}. A few
economic growth models use these data to partially substitute energy for
labor and capital as a motive productive force \citep{AyresWarr2009, Kummel2011}.

The model presented here differs, foremost because civilization is examined
only as a whole, as an evolving organism whose growth is a response only to
its changing ability to access external resources
\citep{Gowdy2013,Herrmann-Pillath2015}. Nothing is said about internal trade
between countries. Neither is a distinction made between human and physical
capital: the capacity to consume external reserves of energy is considered
a complete substitute for both these quantities at global scales. This
enables the model to be strictly thermodynamic, with no requirement for
dimensionally inconsistent fits to prior economic data that are dependent on
the time and place that is considered. The model's validity as an economic
tool rests only on the observation of a fixed ratio between energy
consumption~$a$ and the time integral of inflation-adjusted economic
production~$C$ (Eq.~\ref{eq:alambdaC}). The theoretical interpretation is
that current energy consumption and dissipation sustains all of
civilization's circulations, even human perceptions, insofar as they have
accumulated through prior economic production.

\subsection{Past economic innovation as the engine for current economic growth}

The most direct implication of the existence of a constant value for
$\lambda$~is that economic wealth cannot be decoupled from energy
consumption. For the past, reconstructions of global rates of energy
consumption going back 2000~years are provided in Table~S3 of
\citet{GarrettEF2014}. For the future, CO$_{2}$ emissions will be
inextricably linked to global prosperity for as long as the economy relies on
fossil fuels \citep{GarrettCO2_2009}. Increasing energy efficiency may be
a commonly supposed mechanism for reducing energy consumption while
maintaining wealth. However, as elaborated in Appendix~B, this does not
appear to be the case.


From Eq.~(\ref{eq:alambdaC}), the relative growth rate of civilization wealth~$C$
and its rate of energy consumption~$a$ are equivalent:
\begin{align}
& \textrm{rate}~\textrm{of}~\textrm{return}=\eta=\frac{\textrm{d}\ln{a}}{\textrm{d}t}=\frac{\textrm{d}\ln{C}}{\textrm{d}t}.
\label{eq:eta}
\end{align}
Effectively, like interest on money in the bank, the parameter~$\eta$
represents the ``rate of return'' that civilization enjoys on its
current wealth~$C$, and that it sustains by consuming ever more power.

Since 1970, rates of return for $a$ and $C$ have varied, but both have
averaged $\sim$\,1.90\,\% per year (Fig.~\ref{fig:lambdaplot}).
Substituting Eq.~(\ref{eq:alambdaC}) into Eq.~(\ref{eq:eta}) yields
a relationship between the rate of return and the inflation-adjusted GWP:
\begin{align}
& Y=\eta{C}=\eta\int\limits_{0}^{t}Y(t')\textrm{d}t',
\label{eq:eta_P/C}
\end{align}
or
\begin{align}
& \eta=\frac{Y}{\int\limits_{0}^{t}Y(t')\textrm{d}t'}.
\label{eq:etaYintY}
\end{align}
So, the current rate of return has inertia since it is tied to the past. It
expresses the ratio of current real production to the historical accumulation
of past real production.

The rate of change of civilization's rate of return can be referred to as an
``innovation rate'':
\begin{align}
& \textrm{innovation}~\textrm{rate}=\textrm{d}\ln\eta/\textrm{d}t.
\label{eq:innovationrate}
\end{align}
Referring to an acceleration term d$\ln\eta/$d$t$ as an
innovation might seem a bit arbitrary. However, in Appendix~B it is shown
that it corresponds directly to more traditional economic descriptions of
innovation such as increases in the ``total factor productivity'' or the
``production efficiency''. For example, it is easy to show from
Eqs.~(\ref{eq:alambdaC}) and~(\ref{eq:eta_P/C}) that $\eta$\,$=$\,$\lambda{Y}/a$.
Since $\lambda$~is a constant, it follows that increases in the production
efficiency (or inverse energy intensity) $Y/a$ are equivalent to the
expression for innovation d$\ln\eta/$d$t$. Innovation is
a driving force for economic growth and energy consumption since it follows
directly from Eqs.~(\ref{eq:eta}) and~(\ref{eq:eta_P/C}) that the real GWP
relative growth rate is governed by the relationship
\begin{align}
& \frac{\textrm{d}\ln{Y}}{\textrm{d}t}=\eta+\frac{\textrm{d}\ln\eta}{\textrm{d}t} \label{eq:GDPgrowth} \\
& \textrm{GWP}~\textrm{growth}=\textrm{rate}~\textrm{of}~\textrm{return}+\textrm{innovation}~\textrm{rate}. \nonumber
\end{align}

\begin{figure}[t]\centering
\includegraphics[width=14cm]{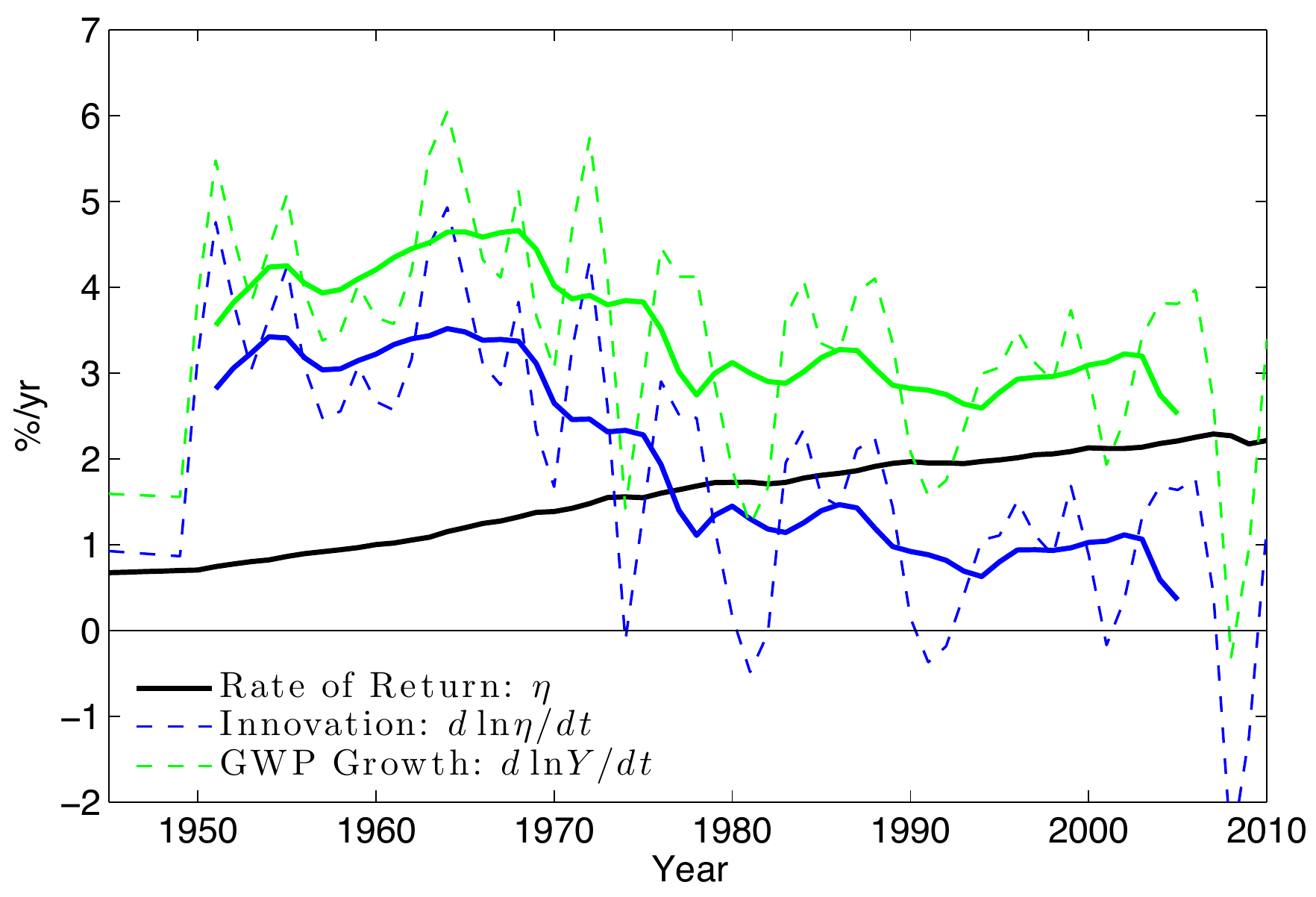}
\caption{Time series of the rate of return, innovation, and the GWP growth
rate, evaluated at global scales and expressed in percent per year. Solid
lines represent a running decadal mean (see \citealp{GarrettEF2014}, for
methods).\label{fig:GDPgrowth}}
\end{figure}

The rate of return~$\eta$ is equivalent to the time integral of past
innovations through $\eta$\,$=$\,$\int\limits_{0}^{t}$\,d$\eta/$d$t'$\,d$t'$, and so
Eq.~(\ref{eq:GDPgrowth}) can be expressed as
\begin{align}
& \frac{\textrm{d}\ln{Y}}{\textrm{d}t}=\int\limits_{0}^{t}(\textrm{d}\eta/\textrm{d}t')\textrm{d}t'+\frac{\textrm{d}\eta/\textrm{d}t}{\int\limits_{0}^{t}(\textrm{d}\eta/\textrm{d}t')\textrm{d}t'}.
\label{eq:dlnYdtaccumulation}
\end{align}
The implication here is that current rates of GWP growth can be considered to
be a consequence of past innovations (the first term) and current innovations
insofar as they are not diluted by past innovations (the second term). The
first term implies that current GWP growth rates will tend to persist because
past innovation is carried to the present; the second term implies that new
technological advances will always struggle to replace older advances that
are already in place \citep{haff2014technology}. Placing an internal
combustion engine on a carriage was revolutionary for its time, but only
a series of more incremental changes have been made to the concept since. Any
new dramatic change has to compete with the large vehicular infrastructure
that has already been put in place.

Figure~\ref{fig:GDPgrowth} shows how rates of return, innovation rates, and
GWP growth have been changing in recent decades based on the data sets
provided in the Supporting Information of Part~1. The rate of return~$\eta$
generally has had an upward trend. In 2008, the rate of return on global
wealth reached an all time historical high of 2.24\,\% per year, up
from 1.93\,\% per year in 1990 and 0.71\,\% per year in 1950.

Meanwhile, innovation rates have declined. The rate of growth of the rate of
return, or d$\ln\eta/$d$t$, has dropped from around 4\,\%
per year in 1950 to near stagnation today. Unprecedented gains in production
efficiency that were obtained in the two decades after World War II
appear to have since given way to much more incremental innovation.

From Eq.~(\ref{eq:GDPgrowth}), GWP growth is the sum of these two pressures.
On the one hand, positive innovation has had a lasting positive impact on the GWP
since it has led to an ever increasing rate of return~$\eta$. On the other
hand, innovation rates have declined. Figure~\ref{fig:GDPgrowth} shows that,
between 1950 and 1970, GWP growth rates were between 4 and 5\,\%
per year. Since 1980, they have been closer to 3\,\% per year.
Increasingly, the long-term increase in civilization's rate of return $\eta$
has been offset by the long-term decrease in innovation
d$\ln\eta/$d$t$. The turning point was in the late 1970s,
when, as shown in Fig.~\ref{fig:GDPgrowth}, innovation rates dipped below
rates of return. Between 1950 and 1975, current innovation was the largest
contributor to current GWP growth rates. Since then, continued GWP growth has
relied increasingly on innovations made in the first two decades since the
end of World War~II (Eq.~\ref{eq:dlnYdtaccumulation}).

\subsection{Physical forces for innovation}

For the special case that there is positive net convergence of matter in a
system, the system grows (see Fig.~\ref{fig:reversible-irreversible} and
Appendix~A). It extends its interface with accessible reserves of energy and
matter. An enlarged interface allows for faster rates of consumption. The
result is a positive feedback that allows growth to accelerate. This is
a basic recipe for emergent or exponential growth. One important aspect of
this feedback, however, is that rates of exponential growth are never
constant. Rather, they increase when new reserves of energy or matter are
discovered and they decrease when there is accelerated decay.

The thermodynamics of this recipe \citep{Garrettmodes2012} were applied in
Part~1 to the emergent growth of civilization and its rates of return on
wealth \citep{GarrettEF2014}. It was shown that the rate of return~$\eta$ can
be broken down into the proportionality
\begin{align}
& \eta\propto(1-\delta)\frac{\Delta{H}_\textrm{R}}{N_\textrm{S}^{2/3}e_\textrm{S}^\textrm{tot}}.
\label{eq:etapropto}
\end{align}
Here, $N_\textrm{S}$ represents the amount of matter or mass within
civilization. $N_\textrm{S}$ grows from a positive imbalance between
civilization's incorporation of raw materials from the environment and
civilization's material decay. $\delta$~is the decay parameter that accounts
for how rapidly $N_\textrm{S}$ falls apart due to natural causes.
$\Delta{H}_\textrm{R}$ represents the size of the energy reserves that are available
to be consumed by civilization. The term $e_\textrm{S}^\textrm{tot}$
represents how much of this energy must be consumed by civilization in order
that raw materials can be added to civilization's fabric, thereby adding to
$N_\textrm{S}$. The exponent~$2/3$ arises from how flows are down
a gradient and across an interface.

Building on the identity $a$\,$=$\,$\lambda{C}$, it was argued in
\citet{GarrettEF2014} that Eq.~(\ref{eq:etapropto}) implies that rates of
economic innovation can be represented by
\begin{align}
& \frac{\textrm{d}\ln\eta}{\textrm{d}t}=-2\eta+\eta_\textrm{tech} \label{eq:innovation_expression} \\
& 
\end{align}
The first term represents a drag on innovation due to a law of diminishing
returns $-$2$\eta$. The second term expresses a rate of technological change
$\eta_\textrm{tech}$ due to changes in $\delta$, $\Delta{H}_\textrm{R}$, and
$e_\textrm{S}^\textrm{tot}$. A distinction is made here between
a technological advance and an innovation. Technological change only counts
as an innovation if it overcomes diminishing returns to lead to a real
increase in the rate of return~$\eta$.

A law of diminishing returns is a characteristic feature of emergent systems.
As indicated by Eq.~(\ref{eq:etapropto}), the exponential growth rates of
larger, older objects with high values of $N_\textrm{S}$ tend to be lower than for
smaller, younger ones. In our case, our bodies are a complex network of
nerves, neurons, veins, gastrointestinal tracts, and pulmonary tubes. We use this
network so that we can interact with a network of electrical circuits,
communication lines, plumbing, roads, shipping lanes, and aviation routes
\citep{Dijk2012}. Such networks have been built from a net accumulation of
matter. So, as civilization grows, any given addition becomes increasingly
incremental.

The implication of Eq.~(\ref{eq:innovation_expression}) is that, absent
sufficiently rapid technological change, relative growth rates~$\eta$ will
tend to decline, and innovation will turn negative. For example, from
Eq.~(\ref{eq:innovation_expression}), innovation requires that
$\eta_\textrm{tech}$\,$>$\,2$\eta$. Or, by substituting Eq.~(\ref{eq:GDPgrowth}) into
Eq.~(\ref{eq:innovation_expression}), an expression for GWP growth is
d$\ln{Y}/$d$t$\,$=$\,$-\eta$\,$+$\,$\eta_\textrm{tech}$, in which case
maintenance of positive GWP growth requires that $\eta_\textrm{tech}$\,$>$\,$\eta$.
That economic growth has been sustained over the past 150~years is
a testament to the importance of technological change for overcoming
diminishing returns.

The rate of technological change follows from the first derivate of Eq.~(\ref{eq:etapropto}):
\begin{align}
& \eta_\textrm{tech}=\frac{\textrm{d}\ln(1-\delta)}{\textrm{d}t}+\frac{\textrm{d}\ln\Delta{H}_\textrm{R}}{\textrm{d}t}-\frac{\textrm{d}\ln{e}_\textrm{S}^\textrm{tot}}{\textrm{d}t} \label{eq:eta_tech} \\
& \eta_\textrm{tech}=\eta_{\delta}+\eta_\textrm{R}^\textrm{net}+\eta_\textrm{e} \nonumber \\
\end{align}
The first of these three forces is improved longevity. Suppose that
civilization decays by losing matter at rate $j_\textrm{d}$. At the same
time, it incorporates new matter at rate $j_\textrm{a}$, as discussed in
Appendix~A. A dimensionless decay parameter
$\delta$\,$=$\,$j_\textrm{d}/j_\textrm{a}$ can be introduced that expresses the
relative importance of material decay to material growth: if $\delta$~is
zero, new material growth is not offset by decay. If $\delta$ declines then
it is because new or existing matter lasts longer, representing an increase
in civilization's longevity. For example, $\delta$ might decrease as
civilization shifts from wood to steel as a construction material. Alternatively, it
might increase due to more frequent natural disasters from climate change.

In Part~1, it was shown how the nominal GWP can be tied to the incorporation
of new matter into civilization $j_\textrm{a}$; the real inflation-adjusted
GWP can be tied to the net incorporation of new matter
$j_\textrm{a}$\,$-$\,$j_\textrm{d}$. This yields the interesting result that
physical decay is related to economic inflation. At domestic scales, the so-called ``GDP
deflator'' is often used as an analog for the annual inflation rate
$\langle{i}\rangle$ since it represents the fractional downward
adjustment that is imposed on the nominal GDP to obtain the real GDP. For
civilization as a whole, the implication is that declining decay, or
increased longevity, corresponds to a smaller GDP deflator, declining
inflation, and faster real GWP growth, i.e.,
\begin{align}
& \eta_{\delta}\simeq-\frac{\textrm{d}\langle\delta\rangle}{\textrm{d}t}\simeq-\frac{\textrm{d}\langle{i}\rangle}{\textrm{d}t}.
\label{eq:eta-delta-inflation}
\end{align}

The second force for technological change in Eq.~(\ref{eq:eta_tech}) is
discovery of new energy reserves. Where discovery exceeds reserve depletion, it
accelerates economic innovation through an increase in the size of available
energy reserves $\Delta{H}_\textrm{R}$ \citep{Smil2006,AyresWarr2009}.
Energy reserves decline as they are consumed at rate~$a$. Meanwhile,
civilization discovers new reserves at rate~$D$. The rate of net discovery is
\begin{align}
& \eta_\textrm{R}^\textrm{net}=\frac{D-a}{\Delta{H}_\textrm{R}}.
\label{eq:discovery}
\end{align}
Provided that reserves expand faster than they are depleted, the rate
$\eta_\textrm{R}^\textrm{net}$ is positive. It represents a technological
advance because there is reduced competition for available resources. From
Eq.~(\ref{eq:etapropto}), larger reserves enable higher rates of return for the
relative growth of wealth and energy consumption.

The specific enthalpy of civilization $e_\textrm{S}^\textrm{tot}$ in
Eq.~(\ref{eq:eta_tech}) is an expression of the amount of power~$a$ that is
required for civilization to extract raw materials and to incorporate them
into civilization's fabric at rate $j_\textrm{a}$. If the ratio
$a/j_\textrm{a}$ declines, then civilization becomes more energy-efficient.

For example, mining and forestry is currently powered by large diesel engines
rather than human and animal labor. Civilization is able to extract raw
materials with comparative efficiency and lengthen civilization networks at
a correspondingly greater rate. Using less energy, we are able to build more
roads, lengthen communications networks, and even increase population, as we
too are made of matter and are part of civilization's fabric. Where the
extraction efficiency of raw materials improves, it is an effective force for
technological change defined by
\begin{align}
& \eta_\textrm{e}=\frac{\textrm{d}\ln{j}_\textrm{a}}{\textrm{d}t}-\frac{\textrm{d}\ln(a)}{\textrm{d}t}.
\label{eq:etae}
\end{align}

\subsection{Deterministic solutions for economic growth}

\begin{figure}[t]\centering
\includegraphics[width=14cm]{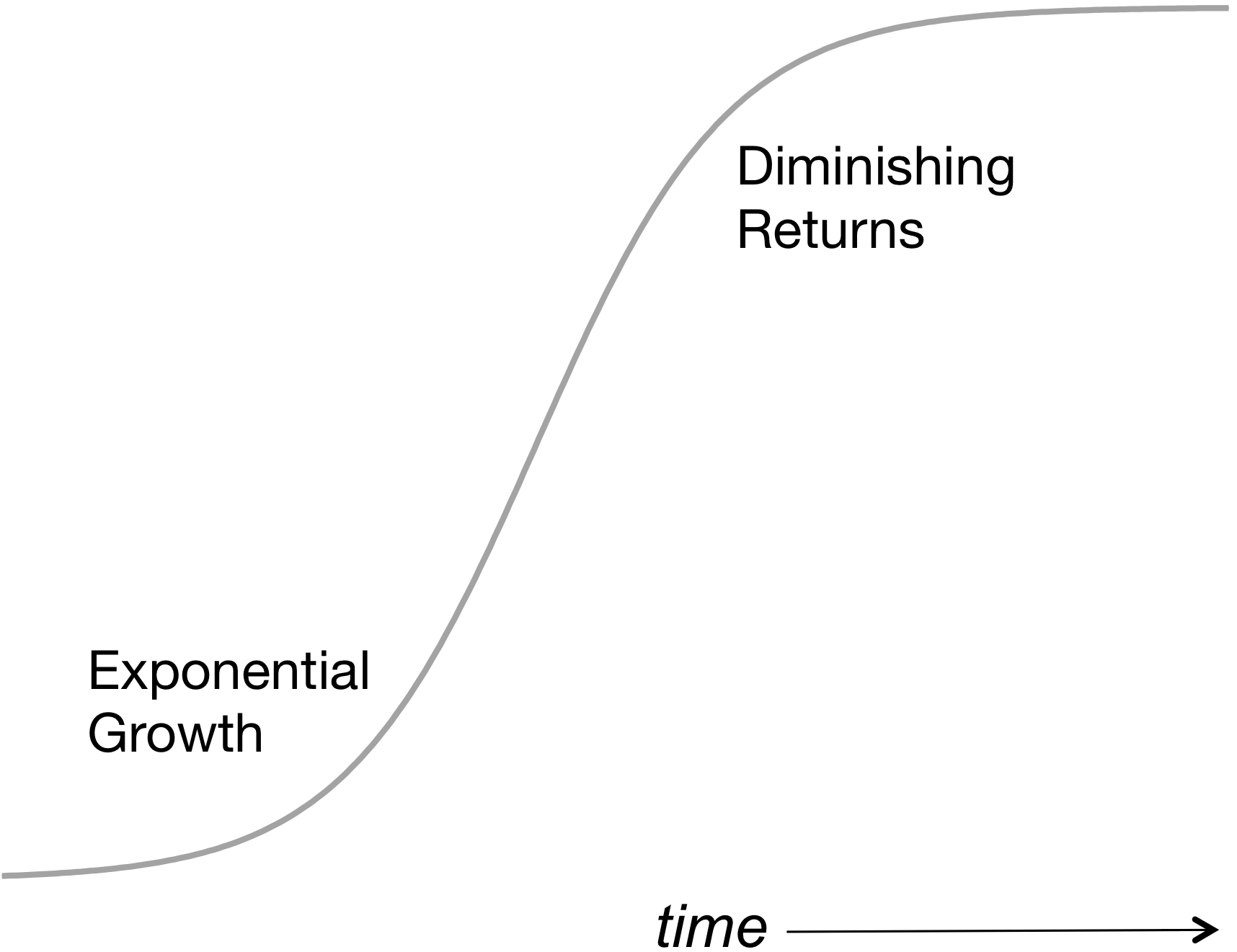}
\caption{Illustration of the logistic curve.\label{fig:Scurve}}
\end{figure}

Equation~(\ref{eq:innovation_expression}) for innovation is logistic in form.
That is, it could be expressed as
d$\eta/$d$t$\,$=$\,$\eta_\textrm{tech}\eta$\,$-$\,2$\eta^{2}$ with a rate of
exponential growth $\eta_\textrm{tech}$ and a drag rate on growth of
$-$2$\eta$. An initial exponential growth phase yields to diminishing returns
where rates of return stabilize (Fig.~\ref{fig:Scurve}). If
$\eta_\textrm{tech}$ is constant, then the solution for the rate of return~$\eta$ is
\begin{align}
&
\eta(t)=\frac{\eta_\textrm{tech}/2}{1+(G-1)\exp\left(-\eta_\textrm{tech}t\right)},
\label{eq:logistic_solution-1}
\end{align}
where
\begin{align}
& G=\frac{1}{2}\frac{\eta_\textrm{tech}}{\eta_{0}}
\label{eq:G-1}
\end{align}
represents a ``growth number'' \citep{Garrettmodes2012,GarrettEF2014} and the
subscript~0 indicates the initial observed value for $\eta_{0}$. The
solution for $\eta$ in Eq.~(\ref{eq:logistic_solution-1}) is sigmoidal.
Provided $G$~is greater than 1, rates of return initially increase
exponentially and saturate at a rate of $\eta_\textrm{tech}/2$. So, for
example, if $\eta_\textrm{tech}$ is sustained at 5\,\% per year, then
one would expect rates of return to grow sigmoidally towards 2.5\,\%
per year. The characteristic time for the exponential growth phase would be
$1/\eta_\textrm{tech}$, or 20~years.

From Eq.~(\ref{eq:GDPgrowth}), the corresponding time-dependent solution for
GWP growth assuming a fixed rate of technological change is
\begin{align}
& \frac{\textrm{d}\ln{Y}}{\textrm{d}t}(t)=\frac{\eta_\textrm{tech}}{2}\left[\frac{1+2(G-1)\exp\left(-\eta_\textrm{tech}t\right)}{1+(G-1)\exp\left(-\eta_\textrm{tech}t\right)}\right].
\label{eq:logistic_solution_GWP}
\end{align}
Here, GWP growth rates also saturate at a value of $\eta_\textrm{tech}/2$,
but if $G$\,$>$\,1 then this is by way of decline rather than growth. Thus, rates
of return on wealth (Eq.~\ref{eq:logistic_solution-1}) and rates of GWP
growth (Eq.~\ref{eq:logistic_solution_GWP}) should have a tendency to
converge with time. This is in fact precisely the behavior that has been
observed in the past few decades. Figure~\ref{fig:GDPgrowth} shows values of
$\eta$ and d$\ln{Y}/$d$t$ that differ by about a factor of
4 in 1950 but that are approaching from opposite directions towards
a common value of about 2.5\,\% per year.

\section{Model validation through hindcasts}

Equation~(\ref{eq:innovation_expression}) is a new expression for the
long-run evolution of the global economy and its resource consumption. Three
approaches are now taken to test its validity.

\subsection{The functional form relating innovation to growth}

\begin{figure}[t]\centering
\includegraphics[width=14cm]{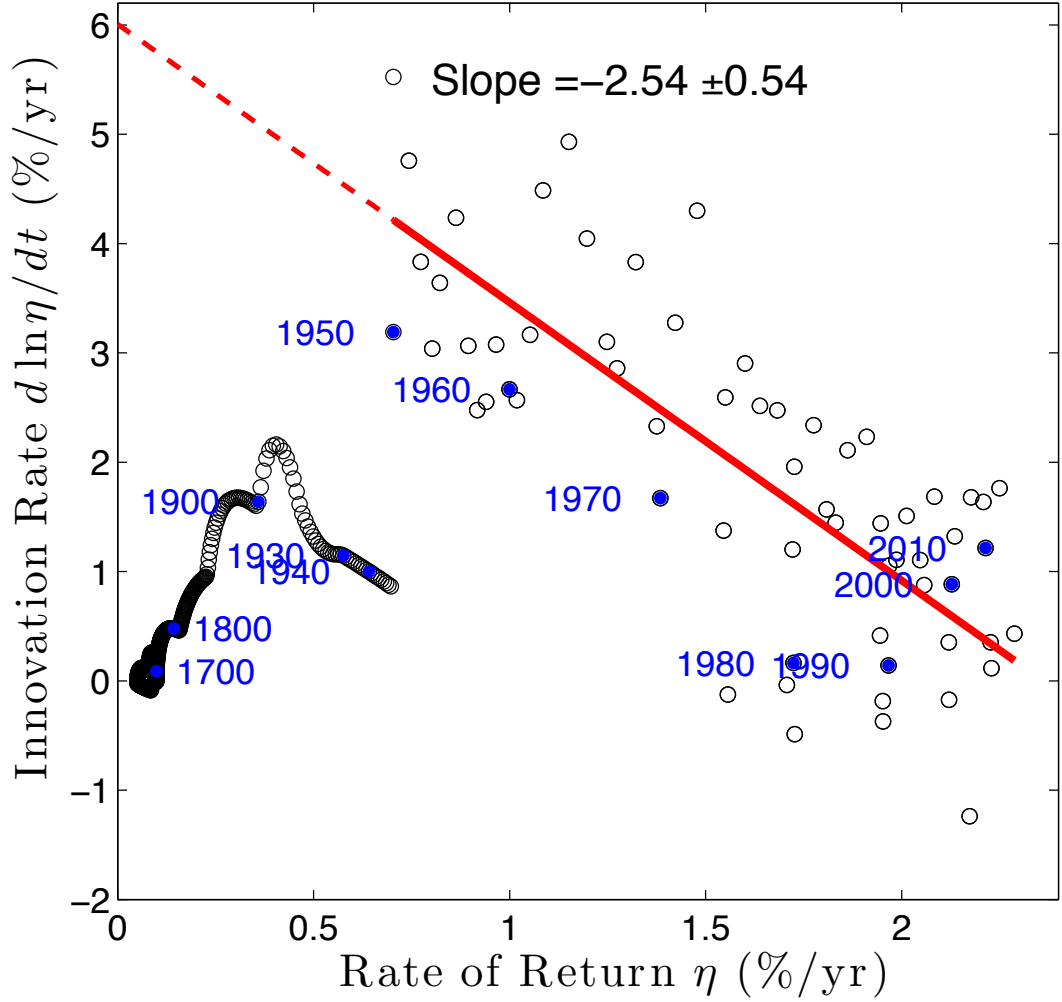}
\caption{The global innovation rate d$\ln\eta/$d$t$ versus
the global rate of return $\eta$\,$=$\,$Y/\int\limits_{0}^{t}\,Y$\,d$t'$
(Eq.~\ref{eq:eta}). Select years are shown for reference. Since 1950,
innovation is related to growth through the functional relationship
d$\ln\eta/$d$t$\,$=$\,$S\eta$\,$+$\,$b$, where the slope and intercept shown
by the red line, with 95\,\% confidence limits, are $S$\,$=$\,$-$2.54\,$\pm$\,0.54
and $b$\,$=$\,0.06\,$\pm$\,0.01.\label{fig:dlnetadt}}
\end{figure}

Figure~\ref{fig:dlnetadt} shows the relationship between innovation rates and
rates of return over the past three centuries derived from GWP estimates from
\citet{Maddison2003} and the United Nations \citep{UNstats}, using Eqs.~(\ref{eq:etaYintY})
and~(\ref{eq:innovationrate}) (see Sect.~2 and the
Supporting Information of Part~1 for methods and associated statistics).
Rapid innovation and accelerating rates of return characterized the
industrial revolution and the late 1940s. Periods of subsiding innovation
followed 1910 and 1950.

Equation~(\ref{eq:innovation_expression}) implies that, if rates of
technological change $\eta_\textrm{tech}$ are roughly a constant, innovation
rates d$\ln\eta/$d$t$ should be related to rates of return~$\eta$
by a slope of about~$-$2; the intercept should be equivalent to the
rate of technological change $\eta_\textrm{tech}$ given by
Eq.~(\ref{eq:eta_tech}). Focussing on the period since 1950, where statistical
reconstructions of GWP are yearly and presumably most reliable
\citep{Maddison2003}, Fig.~\ref{fig:dlnetadt} shows that the past
60~years have been characterized by a least-squares fit relationship
between innovation rates d$\ln\eta/$d$t$ and rates of return~$\eta$
(with 95\,\% uncertainty bounds) given by
\begin{align}
& \frac{\textrm{d}\ln\eta}{\textrm{d}t}=-(2.54\pm0.54)\eta+(0.06\pm0.01).
\label{eq:depletion-observed}
\end{align}
Within the stated uncertainty, the observed slope relating innovation to
rates of return is consistent with the theoretically expected value of~$-$2
that comes from a law of diminishing returns. The implied rate of
technological discovery for this time period $\eta_\textrm{tech}$ is the
intercept of the fit, or about 6\,\% per year. The magnitude of the
difference of the fit from the anticipated slope might be an indication that
$\eta_\textrm{tech}$ has not in fact been a strict constant but rather has declined with slowly with time, as discussed below.


\subsection{Hindcasts of long-run civilization growth}

\begin{figure*}[t]\centering
\includegraphics[width=16cm]{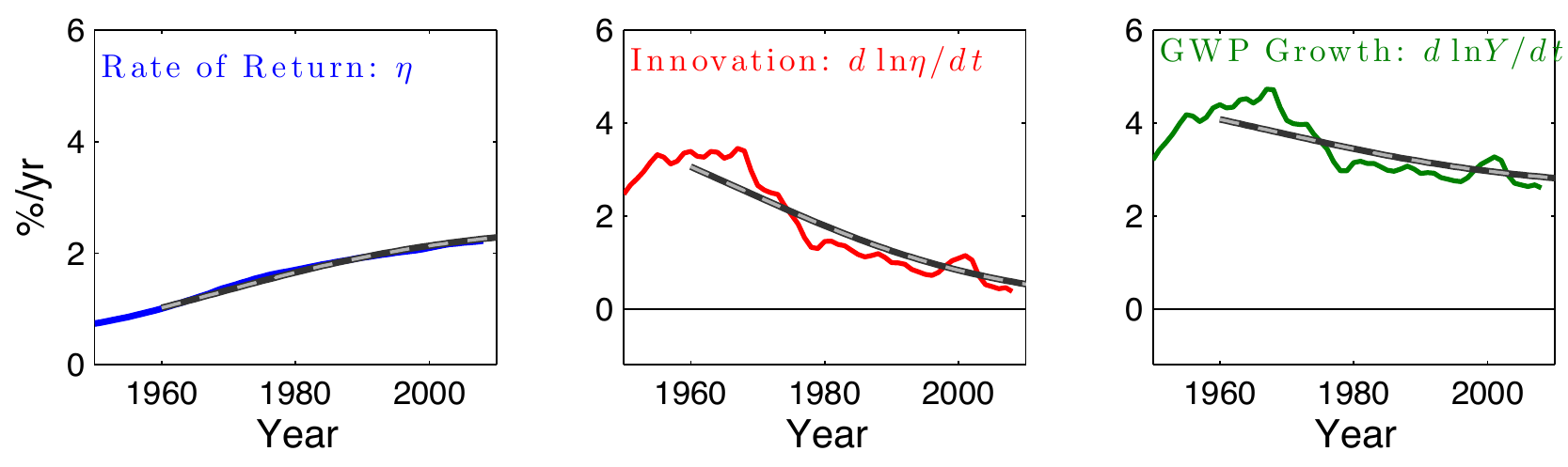}
\caption{Black, gray-dashed lines: hindcasts starting in 1960 of the global rate of return
$\eta$\,$=$\,$Y/\int\limits_{0}^{t}\,Y$\,d$t'$, innovation rates
d$\ln\eta/$d$t$, and the GWP growth rate d$\ln{Y}/$d$t$\,$=$\,$\eta$\,$+$\,d$\ln\eta/$d$t$. Hindcasts are derived from
Eq.~(\ref{eq:logistic_solution-1}) assuming an average rate of technological
change of 5.1\,\%\,yr$^{-1}$ (dashed gray lines) derived from conditions
observed in the 1950s. Solid colored lines: observed decadal running means. Hindcast values in 1960 represent persistence values shown in Table 1. \label{fig:IVP}}
\end{figure*}

The second test is to approach the problem as a hindcast. A hypothetical economic forecaster in
1960 might have noted that the average values of $\eta$ and
d$\ln\eta/$d$t$ between 1950 and 1960 were 0.9\,\% and 3.3\,\% per year, respectively. From
Eq.~(\ref{eq:innovation_expression}), this implies that $\eta_\textrm{tech}$
was 5.1\,\% per year during this period. Applying
Eqs.~(\ref{eq:logistic_solution-1}) and~(\ref{eq:logistic_solution_GWP}), and
using an initial value for $\eta_{0}$ of 1.0\,\% per year in 1960, the
forecaster could then have obtained the trajectories for economic innovation
and growth that are shown in Fig.~\ref{fig:IVP}.

Fifty-year hindcasts are summarized in Table~\ref{tab:hindcasts} along
with skill scores defined relative to a reference model of persistence in
trends \citep{AMS2014}:
\begin{align}
& \textrm{Skill}~\textrm{score}=1-\frac{|\textrm{error(hindcast)}|}{|\textrm{error(persistence)}|}.
\label{eq:skill-score}
\end{align}
Skill scores are positive when the hindcast beats persistence in trends, and
zero when they do not. For example, average rates of energy consumption
growth in the past decade would have been forecast to be 2.3\,\% per
year relative to an observed average of 2.4\,\% per year. Relative to
a persistence prediction of 1.0\,\% per year, the skill score is
96\,\%. Alternatively, a forecast of the GWP growth rate for the first decade of
this century would have been 2.8\,\% per year compared to the actual
observed rate of 2.6\,\% per year. The persistence forecast based on
the 1950 to 1960 period is 4.0\,\% per year, so the skill score is 91\,\%.

\begin{table*}[t]\centering
\caption{For key economic parameters, a comparison between observed annual
growth rates and 50-year predictions made assuming either a reference model
of persistence or a hindcast model given by
Eq.~(\ref{eq:innovation_expression}). Persistence is derived from historical
rates between 1950 and 1960. The ``observed'' time period is 2000 to 2010. The
skill score is derived from 1\,$-$\,error(hindcast)/error(persistence), where
error is derived relative to observed rates. Data are shown in Figure 7. \label{tab:hindcasts}}
\begin{tabular}{lcccc}
\hline & Persistence & Hindcast & Observed & Skill \\
& (\%\,yr$^{-1}$) & (\%\,yr$^{-1}$) & (\%\,yr$^{-1}$) & score \\
& & & & (\%) \\
\hline
Rate of return $\eta$ (d$\ln{a}/$d$t$) & 1.0 & 2.3 & 2.2 (2.4) & 88 (96) \\
Innovation rate d$\ln\eta/$d$t$ & 3.3 & 0.4 & 0.4 & 100 \\
GWP growth rate $\eta$\,$+$\,d$\ln\eta/$d$t$ & 4.0 & 2.8 & 2.6 & 91 \\
\hline
\end{tabular}
\end{table*}

\subsection{Observed magnitude of technological change}

High skill scores suggest that it is possible to provide physically
constrained scenarios for civilization evolution over the coming century
using a simple logistic model given by Eq.~(\ref{eq:logistic_solution-1}). There do not appear to be other macroeconomic
forecast models that are equally successful, and in any case, macroeconomic
models are not normally evaluated through comparisons to multi-decadal
historical data. If a comparison with data is made, it is not in the form of
a true hindcast. The model is judged by the extent to which a sufficiently
complex production function can be tuned to provide an accurate fit to prior
observations \citep[e.g.,][]{WarrAyres2006}.

Using only a fixed value for $\eta_\textrm{tech}$ as input to the model
presented here appears to work very well, at least for global scales. Still,
a more fully deterministic model would not rely on an assumed value for
$\eta_\textrm{tech}$, even if it is a fit to data prior to the date of model
initialization. It is reasonable to anticipate that future rates of resource
discovery and material longevity will evolve with time. Accounting for such
technological change might prove an important consideration for economic and
climate forecasters over the coming century.

To this end, the third test is to try to quantify the thermodynamic
forces outlined in Eq.~(\ref{eq:eta_tech}) that would enable a more first-principles estimation of the value of $\eta_\textrm{tech}$. Methods and
data sets for estimating a time series for the sizes of energy reserves, the
rate of energy consumption, the rate of raw material consumption, and
economic inflation during the period between 1950 and 2010 are described in
Appendix~C and summarized in Table~\ref{tab:Technologicalchangestatistics}. Average rates are shown for
three successive 20-year periods beginning in 1950, and for the 1950
to 2010 period as a whole.

What stands out in Table 2 is how there was unusually rapid technological change between
1950 and 1970. This period was characterized by rapidly growing access to
reserves of oil, gas, and raw materials. It was followed by an abrupt
slowdown in 1970 with no clear long-term recovery since. Summing over these
forces, and averaged over the entire 1950 to 2010 period, rates of
technological change $\eta_\textrm{tech}$ are estimated to have been
a respectable 3.5\,\% per year. Most of this growth took place in the
first 20~years, when it achieved 7.0\,\% per year. The latest
20-year period averaged just 1.4\,\% per year.

\begin{table*}[t]\centering
\caption{Components of technological change expressed as 20- and 60-year
averages of growth rates. Bold numbers represent weighted averages. See Section~3.3 and Appendix~B for details.
\label{tab:Technologicalchangestatistics} }
\begin{tabular}{lllll}
\hline
Mean growth rates (\%\,yr$^{-1}$) & 1950--1970 & 1970--1990 & 1990--2010 & 1950--2010 \\
\hline
\textbf{Average raw materials per energy} $\boldsymbol{\eta}_\mathbf{e}$ & \textbf{3.5} & \textbf{--0.7} & \textbf{0.7} & \textbf{1.3} \\
Cement and wood per energy & 2.2 & $-$0.8 & $-$0.4 & 0.5 \\
Iron and steel per energy & 4.6 & $-$1.4 & 1.4 & 1.7 \\
Copper per energy & 3.7 & 0.0 & 1.0 & 1.6 \\
\hline
\textbf{Total fossil reserves} $\boldsymbol{\eta}_\mathbf{R}^\mathbf{net}$ & \textbf{3.6} & \textbf{1.3} & \textbf{0.7} & \textbf{2.0} \\
Oil reserves {[}{production in EJ$/$year}{]} & 3.6 {[}{59}{]} & 0.6 {[}{133}{]} & $-$0.7 {[}{165}{]} & 1.1 {[}{118}{]} \\
Gas reserves {[}{production in EJ$/$year}{]} & 8.2 {[}{22}{]} & 2.4 {[}{62}{]} & 0.6 {[}{98}{]} & 3.7 {[}{60}{]} \\
Coal production {[}{production in EJ$/$year}{]} & 2.2 {[}{73}{]} & 1.9 {[}{115}{]} & 2.3 {[}{153}{]} & 2.2 {[}{113}{]} \\
\hline
\textbf{Change in longevity} $\boldsymbol{\eta_{\delta}}$ & \textbf{--0.1} & \textbf{0.2} & \textbf{0.2} & \textbf{0.2} \\
\hline
\textbf{Rate of technological change} $\boldsymbol{\eta}_\textbf{tech}$\,$=$\,$\boldsymbol{\eta}_\mathbf{e}$\,$+$\,$\boldsymbol{\eta}_\mathbf{R}^\mathbf{net}$\,$+$\,$\boldsymbol{\eta_{\delta}}$ & \textbf{7.0} & \textbf{0.8} & \textbf{1.4} & \textbf{3.5} \\
\hline
\end{tabular}
\end{table*}

\begin{table*}[t]\centering
\caption{Twenty- and 60-year averages of rates of return (calculated using two
independent techniques), innovation rates, and rates of technological change.
Values are derived from Eqs.~(\ref{eq:innovation_expression})
and~(\ref{eq:eta_tech}) and using data from
Table~\ref{tab:Technologicalchangestatistics}.\label{tab:Economicgrowthrates}}
\begin{tabular}{lcccc}
\hline
Mean growth rates (\%\,yr$^{-1}$) & 1950--1970 & 1970--1990 & 1990--2010 & 1950--2010 \\
\hline
Observed rate of return $\eta$\,$=$\,$Y/\int\limits_{0}^{t}\,Y$d$t'$ ($=$\,(d$a/$d$t)/a$) & 1.0 & 1.7 (1.6) & 2.1 (2.0) & 1.6 \\
Observed innovation rate d$\ln\eta/$d$t$ & 3.3 & 1.6 & 0.6 & 1.9 \\
Calculated technological change $\eta_\textrm{tech}$\,$=$\,d$\ln\eta/$d$t$\,$+$\,2$\eta$ & 5.3 & 5.0 & 4.7 & 5.1 \\
Observed technological change $\eta_\textrm{tech}$ & 7.1 & 0.8 & 1.2 & 3.5 \\
\hline
\end{tabular}
\end{table*}

Improved access to energy reserves and raw materials explains most of the
variability in $\eta_\textrm{tech}$. Coal power production expanded steadily
at a rate of about 2\,\% per year. Oil reserves, on the other hand,
expanded at an average 3.6\,\% per year between 1950 and 1970 but
shrunk at an average 0.7\,\% per year between 1990 and 2010. The
amount of energy required to access key raw materials such as cement, wood,
copper, and steel dropped by an average 3.5\,\% per year between 1950
and 1970, which implies rapid efficiency gains. Since 1970, energy
consumption and raw material consumption have grown at nearly equivalent rates,
implying no associated force for technological change.

As a check on the first-principles estimate that the average value of
$\eta_\textrm{tech}$ between 1950 and 2010 was 3.5\,\% per year,
Table~\ref{tab:Economicgrowthrates} shows the 20- and 60-year averages
of $\eta$ and d$\ln\eta/$d$t$, and it uses these to derive a
rate of technological change $\eta_\textrm{tech}$ from
Eq.~(\ref{eq:innovation_expression}). (As a consequence of $\lambda$ being
a constant, calculated rates of return~$\eta$ are similar whether they are
calculated from available energy statistics using Eq.~(\ref{eq:eta}) or from
GWP statistics using Eq.~(\ref{eq:etaYintY}). Both have averaged
1.6\,\% per year overall.)

Innovation rates d$\ln\eta/$d$t$ have been positive overall,
meaning rising rates of return. Still, they declined from 3.3\,\% per
year between 1950 and 1970 to just 0.6\,\% per year between 1970 and
1990. Thus, the estimated average rate of technological change derived from
Eq.~(\ref{eq:innovation_expression}) (i.e., $\eta_\textrm{tech}$\,$=$\,d$\ln\eta/$d$t$\,$+$\,2$\eta$) is 5.1\,\%
per year, similar to what was derived for the 1950 to 1960 time period as
discussed in Sect.~3.2. In comparison, the rate of technological change
estimated from the physical parameters described in
Table~\ref{tab:Technologicalchangestatistics} averages 3.5\,\% per
year, or about one-third lower. Whether the residual 1.6\,\% per year
is due to data uncertainties or theoretical considerations is unknown.

The hindcasts in Sect.~3.2 assumed a constant value for $\eta_\textrm{tech}$,
whereas the observed rates summarized in Table~\ref{tab:Economicgrowthrates}
point towards much higher variability. Perhaps the reason a constant value
nonetheless leads to hindcasts with high skill scores is because there is
a timescale of decades for externally forced technological change to diffuse
throughout the global economy \citep[e.g.,][]{rogers2010diffusion}. Assuming
a fixed value for $\eta_\textrm{tech}$ represents this timescale by smoothing
the economy-wide impacts of a large impulse of innovative forces that
occurred between 1950 and 1970.

\section{Positive skill in economic forecasts}

A logistic equation forms the basis of the prognostic model 
that provides hindcasts for civilization growth as a whole. At the level of empires, there have been similar
waves of logistic or sigmoidal growth throughout history. An initial phase of
exponential growth tends to be followed by slower rates of expansion. Ancient
Rome's empire increased to cover 3\,500\,000\,km$^{2}$ in its first
300~years, but only a further 1\,000\,000\,km$^{2}$ in its
second; the Mongol empire extended to 20\,000\,000\,km$^{2}$ within
50~years, adding an additional 4\,000\,000\,km$^{2}$ in the next
\citep{Marchetti2012}. Growth at declining rates has also been noted in the
adoption of new technologies \citep{rogers2010diffusion}, the size of oil
tankers \citep{Smil2006}, bacteria \citep{zwietering1990modeling}, and
snowflakes \citep{PruppacherKlett1997}.

In a very general way, these common emergent behaviors might be viewed as the
response of a system to available reserves of potential energy and matter.
Consumption of resources allows for expansion into more resources and hence
more consumption. Eventually, new consumption becomes increasingly diluted by
past consumption in which case growth slows. The mathematical expression of
the dynamics is fairly simple \citep{GarrettEF2014}, and it has been shown
here how it can serve as a foundation for making 50-year hindcasts of
the global economy (Fig.~\ref{fig:IVP}).

\begin{figure}
\includegraphics[width=14cm]{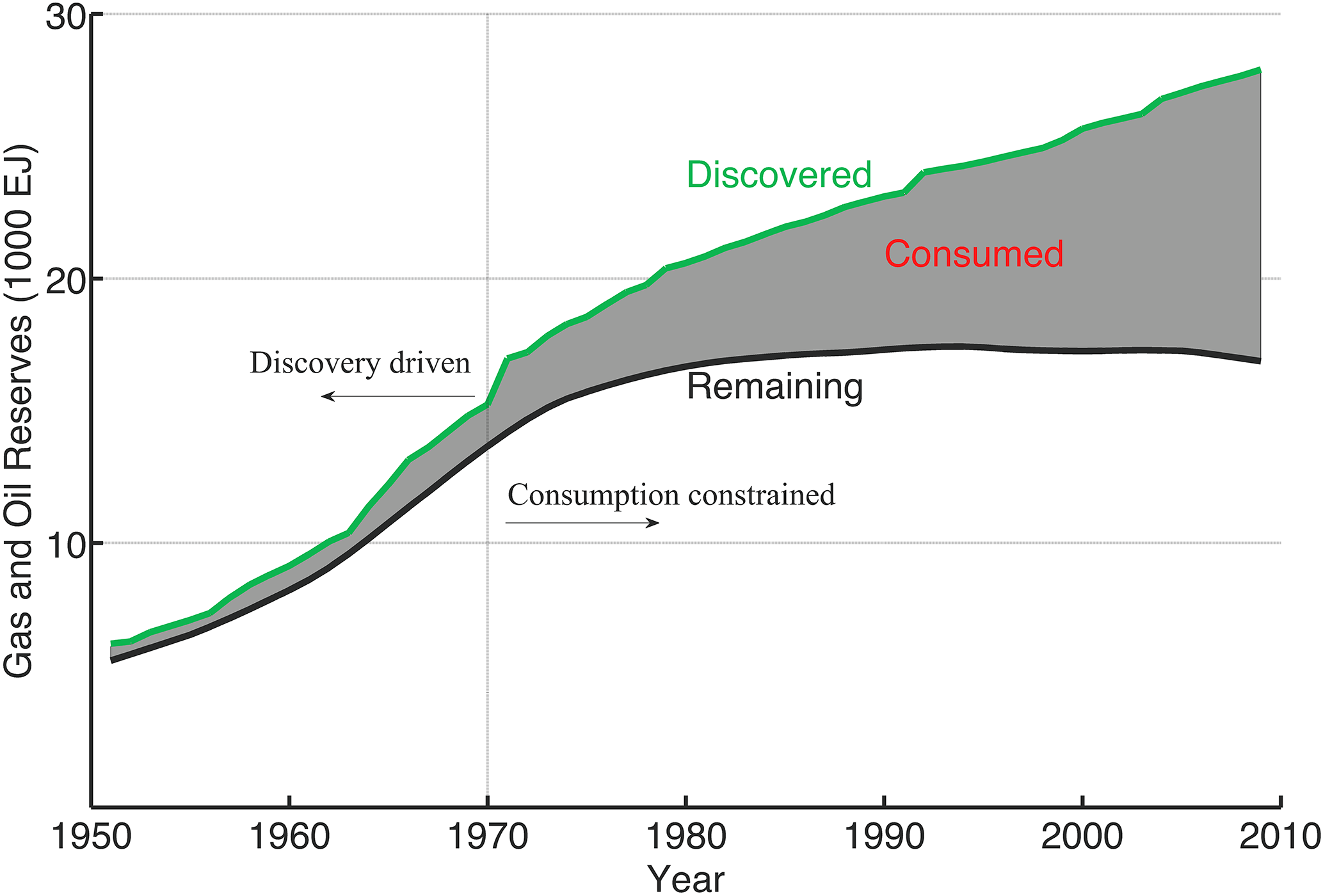}
\caption{Discovered, consumed, and remaining global reserves of gas and oil
since 1950 \citep[source:][]{IHS}.\label{fig:gasandoil}}
\end{figure}

The accuracy of the hindcasts shown here appears to be due in part to a remarkable
burst of technological change that occurred between 1950 and 1970.
Figure~\ref{fig:gasandoil} encapsulates its magnitude. From available
statistics, oil and gas reserves expanded faster than they were consumed.
This changed around 1970. Reserves continued to be uncovered, but they only
barely kept pace with increasing demand. Early innovation and growth began to
act as a drag on future innovation.

How rapid resource discovery played out is captured mathematically by
Eq.~(\ref{eq:innovation_expression}), at least assuming a fixed value for
$\eta_\textrm{tech}$ of about 5.1\,\% per year. Forecasting future scenarios may not be so easy because
the evolutionary behavior is most clear when $\eta_\textrm{tech}$ is large.
Civilization growth rates~$\eta$ have nearly completed their adjustment to
the asymptotic value of $\eta_\textrm{tech}/2$ that is predicted by
Eq.~(\ref{eq:logistic_solution-1}). What this implies is that, because
innovation appears to have dropped to relatively low levels in recent
decades, there is no longer a clear past signal that can be relied upon to
propel civilization forward in a prognostic model; the post-war impulse has
largely run its course.

This does not mean that the model described here lacks utility looking
forward; rather, it implies that $\eta_\textrm{tech}$ must be derived from
something more than a fit to the past. To this end, three forces for
technological change were identified (Eq.~\ref{eq:eta_tech}). One is how fast
civilization networks fray from such externalities as natural disasters. The
others address the accessibility of raw materials and how fast new energy
reserves are discovered relative to their rates of depletion. Predictions of
how these three factors combine may provide a basis for future scenarios for
humanity, based more on external physical forces than internal human policies.

\section{Conclusions}

In Lewis Carroll's \textit{Through the Looking Glass}, Alice was urged by the
Red Queen to run with her ever faster. But, ``however fast they went, they
never seemed to pass anything''. As the Red Queen put it, ``Now,
\textit{here}, you see, it takes all the running you can do, to keep in the
same place''. In the 1950s and 1960s, civilization made exceptionally rapid
gains in energy reserve discovery and resource extraction efficiency. This
spurred a rapid acceleration of growth in global wealth that required an
equal demand for energy. What followed post-1970 was more constrained growth
because diminishing returns settles in for any large system and because
fossil fuel resource discovery only just kept up with increasing demand \citep{Bardi2009,Murray2012}.


Further along, we might anticipate that decay from natural disasters and
environmental degradation will also play an important role in civilization's growth trajectory
\citep{arrow1995economic}. Statistics presented here suggest that decay has
thus far been a comparatively weak player. This may change if, as expected, atmospheric CO$_{2}$
concentrations reach ``dangerous'' levels and decay rates increase
\citep{HansenDangerous2007,Matthews2009,Garrettcoupled2011,Mora2013}.

Should diminishing returns, resource depletion, and decay combine to cause
civilization growth to stall, then simulations described in Part~1 suggest
that external forces may have the potential to push civilization into a phase
of accelerating decline. Civilization lacks the extra energy required to
compensate for continued natural disasters, much less grow, and so it tips
towards collapse.

Contraction of wealth implies a rate of return~$\eta$ that is negative
(Eq.~\ref{eq:eta}). From Eq.~(\ref{eq:eta_P/C}), this suggests a global economy with
a positive nominal GWP but, in effect, a negative real GWP. Fortunately,
recent history does not provide a guide for such a global economic disaster.
Still, one might imagine a scenario where historically accumulated global
wealth shrinks because, at regional or sectoral levels, an ever smaller
fraction of civilization remains involved in gross economic production. A
nominal GWP remains to be tallied, but it is increasingly offset elsewhere by
some combination of wars, a degrading environment, growing unemployment,
inflation, death, and decay. Energy consumption is still required to support
society -- after all, we must always eat. But a diminishing portion
of society is able to add net value calculated with monetary instruments that
offer promises of future returns.

The silver lining of a contracting civilization might be slowing
CO$_{2}$ emissions, and eventually slower climate change. The model
introduced in Part~1 for making multi-decadal hindcasts of civilization
evolution allows for both positive and negative feedbacks to be represented
in the coupled evolution of the human--climate system. This paper shows that
the economic side of this model is successful at reproducing the past
50~years of economic growth. The next step will be to use the model to
provide a range of physically constrained forecasts for the evolution of
civilization and the atmosphere for the remainder of this century.


\appendix
\section{Reversible cycles and irreversible flows}

An implicit consideration with the approach taken here is that it separates
small, short-term, ``micro-'' economic behaviors from larger, longer-term,
``macro-'' economic evolution. From the perspective of thermodynamics,
short-term equilibrium, reversible, cyclic behaviors that are not explicitly
resolved are separated from longer-term non-equilibrium, irreversible dynamics
that are resolved. This is a common strategy, one illustrated in
Fig.~\ref{fig:reversible-irreversible}, as familiar as the separation of the
tachometer and speedometer in a car. One represents reversible engine cycles,
whereas the other expresses the rate of irreversible travel down the road.

In general, reversible and irreversible processes are linked. This is because
the second law of thermodynamics prescribes that all processes are irreversible. Introducing
the concept of reversible circulations within a system is a useful
idealization. However, such circulations can only be sustained by an
external, irreversible flow of energy and matter through the system. When
open systems are near a balance or a steady state, then reversible
circulations can be represented as a four-step Carnot cycle or heat engine
whereby external heating raises the system potential so that raw materials
diffuse from outside the system to inside the system \citep{Zemanksy1997}.
Waste heat is dissipated to the environment so that the system can relax to
its ground potential state where it releases exhaust or undergoes decay.
Averaged over time, the circulations within the system maintain a fixed
amplitude and period $\tau_\textrm{circ}$.

For example, the dynamic circulations of a hurricane are sustained by
a inflow of oceanic heat and an outflow of thermal radiation to space
\citep{Emanuel1987}. In the case of civilization, we consume energy in order
to sustain circulations and extract raw materials from the environment,
leaving behind material waste and radiated heat. Petroleum in a car propels
our material selves to and from work, where we consume carbohydrates,
proteins,
and fats to propel electrical signals to and from our brains so that we can
consume electricity from coal in order to propel charge along copper wires to
and from our computers. Through radiation, frictional losses, and other
inefficiencies, all potential energy is ultimately dissipated as waste heat
to the atmosphere and ultimately through radiation to space at the mean
planetary blackbody temperature of 255\,K. Over short timescales,
consumption approximately equals dissipation, and civilization circulations
maintain a steady state.

Over longer timescales, any small imbalance between consumption and
dissipation by civilization becomes magnified. Raw materials are slowly
incorporated into civilization's fabric at rate $1/\tau_\textrm{growth}$
(Fig.~\ref{fig:reversible-irreversible}). Further, at the same time that
civilization grows, resources are discovered and depleted, and perhaps as
a consequence of climate change, decay rates increase too. The focus shifts
from the short-term reversible circulations associated with our daily lives
to the longer-term timescales associated with the non-equilibrium,
irreversible growth of civilization as a whole.

\section{Comparisons with traditional economic frameworks}

The definition for innovation d$\ln\eta/$d$t$ that has been
introduced here is very similar to definitions that have been made elsewhere.
Traditional neoclassical growth models calculate the nominal growth of
``physical capital''~$K$ (in units of currency) from the difference between the
portion~$s$ of production $P$ that is a savings or investment and capital
depreciation at rate~$\delta$
\begin{align}
& \frac{\textrm{d}K}{\textrm{d}t}=(Y-W)-\delta{K}=sY-\delta{K},
\label{eq:dasKapital}
\end{align}
where individual and government consumption is represented by
$W$\,$=$\,(1\,$-$\,$s)Y$. What is not saved or invested in the future is
consumed in the present.

Labor~$L$ (in units of worker hours) uses accumulated investments in physical
capital to enable further production $Y$ according to some functional form
$f(K$, $L)$. A commonly used representation is the Cobb--Douglas
production function
\begin{align}
& Y=AK^{\alpha}L^{1{-}\alpha},
\label{eq:Cobb-Douglas}
\end{align}
where $A$ is a ``total factor productivity'' that accounts for any residual
in the output~$Y$ that is not explained by the inputs~$K$ and~$L$. The
exponent~$\alpha$ is empirically determined from a fit to past data and
$\alpha$\,$\neq$\,1. Unfortunately, this presents the drawback that the units for
$A$ are ill-defined and dependent on the scenario considered.

The Solow growth model \citep{Solow1957} expresses the prognostic form for
Eq.~(\ref{eq:Cobb-Douglas}) as
\begin{align}
& \frac{\textrm{d}\ln{Y}}{\textrm{d}t}=\frac{\textrm{d}\ln{A}}{\textrm{d}t}+\alpha\frac{\textrm{d}\ln{K}}{\textrm{d}t}+(1-\alpha)\frac{\textrm{d}\ln{L}}{\textrm{d}t}.
\label{eq:Solow}
\end{align}
The term d$\ln{A}/$d$t$ has often been interpreted to
represent technological progress. Such progress might be exogenous
\citep{Solow1957} or endogenous \citep{Grossman1990,Romer1994}. If exogenous,
then progress is considered to be due to an unknown external force. If
endogenous, then it might come from targeted investments such as research and development.

In comparison, the alternative approach that has been presented here
considers civilization as a whole. Labor is subsumed into total capital. The
physics of energy dissipation suggest a focus on the connections between
components within a global network rather than on the elements themselves.
This adjustment requires only a slight, though important, modification to the Solow
growth model in which $\alpha$\,$=$\,1 and $Y$\,$=$\,$A\,K$, where $A$ has fixed units of
inverse time. In this case, no fit to data is required to obtain $\alpha$,
and the units for $A$ are consistent and physical. Equation~(\ref{eq:Solow})
becomes equivalent to the expression $Y$\,$=$\,$\eta{C}$ in
Eq.~(\ref{eq:eta_P/C}),
where $\eta\equiv A$ and $C\equiv K$. Further, the expression d$\ln{A}/$d$t$
that is assumed to describe technological progress in
neoclassical frameworks (Eq.~\ref{eq:Solow}) is then mathematically
equivalent to the definition for innovation d$\ln\eta/$d$t$
in the thermodynamic framework (Eq.~\ref{eq:innovationrate}).

In energy economics, the term ``production efficiency'', or its inverse, the
``energy intensity'', is often used to relate the amount of economic output
that society is able to obtain per unit of energy it consumes
\citep{Sorrell_UKERC2007}. More efficient, less energy-intense production is
ascribed to technological change \citep[e.g.,][]{Pielke2008}.

The production efficiency can be defined mathematically as the ratio
\begin{align}
& f=Y/a.
\label{eq:f}
\end{align}
From Eq.~(\ref{eq:f}) and the expression $a$\,$=$\,$\lambda{C}$
(Eq.~\ref{eq:alambdaC}), where $\lambda$~is a constant, the production
efficiency can then be linked to wealth~$C$ through
\begin{align}
& f=\frac{1}{\lambda}\frac{Y}{C}.
\label{eq:P}
\end{align}
With rearrangement, $\lambda{f}$~is then equivalent to the rate of return~$\eta$ in
Eq.~(\ref{eq:eta_P/C}) that expresses how fast economic wealth~$C$ can be
converted to economic production~$Y$ through $Y$\,$=$\,$\eta{C}$. More efficient
production leads to faster growth of wealth through
\begin{align}
& \eta=\frac{\textrm{d}\ln{C}}{\textrm{d}t}=\lambda{f}.
\label{eq:eta-1}
\end{align}
It follows that
\begin{align*}
& \frac{\textrm{d}\ln\eta}{\textrm{d}t}\equiv\frac{\textrm{d}\ln{f}}{\textrm{d}t}.
\end{align*}
Increasing energy efficiency equates with innovation as
defined by Eq.~(\ref{eq:innovationrate}).

As a side note, since $\eta$~is also equal to the rate of growth in energy
consumption (Eq.~\ref{eq:eta}), this yields the counterintuitive result that
higher production efficiency accelerates growth in energy consumption. What
is normally assumed is the reverse \citep{PacalaSocolow2004,Raupach2007}.
While the concept of ``backfire'' has been reached within more traditional
economic contexts \citep{Saunders2000,Alcott2005}, it is a conclusion that
remains highly disputed, at least where economies are viewed at purely
sectoral levels \citep{Sorrell_UKERC2007,Sorrell2014}.

Here, increased production efficiency $f$\,$=$\,$Y/a$ leads to an acceleration of
energy consumption at rate $\eta$\,$=$\,$\lambda{f}$ because it expands civilization's
boundaries with new and existing energy reservoirs \citep{GarrettEF2014}.
Energy reservoirs may eventually be depleted, but at any given point in time efficiency
permits the positive feedback that leads to ever faster rates of consumption.


A simple example is to contrast a sick child with a healthy child. Without
having to know the ``sectoral'' level details of cellular function, it is
clear that a healthy child will grow fastest. Health here is an implicit
representation of the child's ability to efficiently convert current food
consumption to growth and increased future consumption. Food contains the
energy and matter that the child requires to grow to adulthood. At this
point, hopefully, a law of diminishing returns takes over so that weight is
able to maintain a steady state.

\section{Estimated rates of technological change}

Estimates of technological change rates $\eta_\textrm{tech}$ require global-scale statistics for the size of energy reserves, the rate of energy
consumption, the rate of raw material consumption, and economic inflation.
A challenge is that the reliability and availability of statistics diminishes
the further back one goes in time. Accurate record keeping can be a challenge
even for the most developed nations, much less for every nation. While global
statistics for inflation might be available since 1970, they are given for
only for a few countries in the 1950s \citep[United][]{UNstats}.

It is also not obvious how to sensibly represent raw material consumption.
Cement, steel, copper, and wood may be among the more obviously important
components of the material flow to civilization, but their proportionate
weights are far from clear. Steel is consumed in much greater volume than
copper since it is a basic building material. But copper is an efficient
conduit for electricity and equally important for civilization development.

With respect to energy reserves, the focus here is on fossil fuels since they
remain the primary component of the global energy supply. Energy resources
represent a total that may ultimately prove recoverable. Energy reserves
represent the fraction of resources that is considered currently accessible
given existing political and technological considerations. Unfortunately,
there is no precise definition of what this means. Moreover, reserve and
resource estimates are provided by countries and companies that may have
political reasons to misrepresent the numbers \citep{Hook2010,Sorrell2010}.

The thermodynamic term $\Delta{H}_\textrm{R}$ in Eq.~(\ref{eq:discovery})
represents the potential energy that is available to drive civilization
flows. It would seem to be most obviously represented by reserves rather than
resources since reserves are what are most accessible and they most directly
exert an external pressure on civilization. A question that arises is how to
provide some self-consistent way to add reserves of solid coal to reserves of
natural gas and oil that diffuse to civilization as a fluid.
Thermodynamically, any form of fossil fuel extraction requires some energy
barrier to be crossed, or an amount of work that must be done, in order to
make the potential energy immediately available so that it can diffuse to the
economy. The rate of diffusion is proportional to a pressure gradient (in units
energy density).

For example, well pressure forces a fluid fuel to the surface. Once the
energy barrier of building the well is crossed, the magnitude of the pressure
can be related to the well reserve size $\Delta{H}_\textrm{R}$
\citep{Hook2014}. In contrast, coal reserves must be actively mined with
a continuous energy expenditure. Even if the coal reserve is discovered,
there remains a clear energetic cost in order to obtain an energetic return
\citep{Murphy2010,kiefer2013energy}. A hint at the importance of this energy
barrier is that new fluid fuel reserves like oil and gas appear to affect
economies much more rapidly than coal \citep{Bernanke1997,Stijns2005,Hook2010,Hook2014}, perhaps because they are
more easily extracted and consumed.

In what is hopefully a defensible first step, the aforementioned concerns are
addressed as follows for the purpose of calculating rates of technological
change $\eta_\textrm{tech}$. Rates of growth of energy reserves
(Eq.~\ref{eq:discovery}) are determined assuming that coal consumption is not
reserve-constrained, and rather that the closest solid equivalent to reserves
of oil and gas in terms of accessibility is coal-fired power plants. Like
discovering and exploiting an oil well, a power plant must be constructed,
and it is only at this point that the coal reserve can be accessed to power
civilization. Total reserves are then the production weighted sum of the
rates of growth of coal production capacity, oil reserves, and gas reserves \citep{Rutledge2011,IHS}.


Changes in civilization longevity are estimated using
Eq.~(\ref{eq:eta-delta-inflation}), which expresses decay in terms of
inflation. Global inflation statistics since 1970 are readily available
\citep[United][]{UNstats}. For the period before 1970, an estimate is an
average is taken of the respective inflation rates from the USA, Great
Britain, Japan, Germany, Italy, and France \citep{Inflationstats}.

Rates of change in the specific energy of raw material extraction
$e_\textrm{s}^\textrm{tot}$\,$=$\,$a/j_\textrm{a}$ (Eq.~\ref{eq:etae}) are
derived from statistics for global rates of energy consumption from all
sources~$a$ \citep{AER2011}, and from statistics for the consumption of iron
and steel, copper, wood (excluding fuelwood), and cement
\citep{FAOwoodstats2012,CDIAC_CO2emissions2013,Copperstats2014,Ironandsteelstats2014}.
Wood and cement are treated as substitutable construction materials and are
added according to their respective volumes. The total rate of change in
$j_\textrm{a}$ is then a simple average of the three rates of change: wood
and cement, copper, and iron and steel.

Statistics for the components of technological change are provided in
Table~\ref{tab:Technologicalchangestatistics}.


\section*{Acknowledgements}
This work was supported by the Kauffman Foundation, whose views it does not
claim to represent. Statistics used in this study are available freely
through referenced sources. Andrew~Jarvis, Peter~Haff,
Carsten~Herrmann-Pillath, and Chris~Garrett are thanked for their
constructive reviews of the manuscript. 

\bibliographystyle{agu08}
\bibliography{References}

\end{document}